\documentclass[10pt,conference]{IEEEtran}

\usepackage{amsmath,amsfonts}
\usepackage{dsfont}
\usepackage{graphicx}
\usepackage[hang]{subfigure}
\usepackage{citesort}
\usepackage{color}

\newtheorem{definition}{Definition}
\newtheorem{remark}{Remark}
\newtheorem{theorem}{Theorem}

\newtheorem{lemma}{Lemma}
\newtheorem{corollary}{Corollary}

\newcommand{\ave}{\mu^{(n)}}
\newcommand{\An}{A_{\epsilon}^{(n)}}

\newcommand{\prob}{\mathbb{P}}									%probability
\newcommand{\N}{\mathbb{N}}
\newcommand{\R}{\mathbb{R}}

\newcommand{\sJ}{\mathcal{J}} 
\newcommand{\sK}{\mathcal{K}}
\newcommand{\sL}{\mathcal{L}}
\newcommand{\sM}{\mathcal{M}} \newcommand{\rM}{\mathrm{M}}
 
\newcommand{\sP}{\mathcal{P}}
\newcommand{\sQ}{\mathcal{Q}} \newcommand{\rQ}{\mathrm{Q}}
\newcommand{\sR}{\mathcal{R}}

\newcommand{\sU}{\mathcal{U}} \newcommand{\rU}{\mathrm{U}}
\newcommand{\sV}{\mathcal{V}} \newcommand{\rV}{\mathrm{V}}

\newcommand{\sX}{\mathcal{X}}	\newcommand{\rX}{\mathrm{X}}
\newcommand{\sY}{\mathcal{Y}} \newcommand{\rY}{\mathrm{Y}}

\newcommand{\sCbbc}{\mathcal{C}_{\text{BBC}}}

\newcommand{\Ro}{R_1}
\newcommand{\Rt}{R_2}

\newcommand{\Ri}{R_i}
\newcommand{\Rc}{R_c}
\newcommand{\Requi}{R_e}

\newcommand{\Mon}{M_1^{(n)}}
\newcommand{\Mtn}{M_2^{(n)}}

\newcommand{\Min}{M_i^{(n)}}
\newcommand{\Mcn}{M_{c}^{(n)}}

\newcommand{\tn}{^{\otimes n}}

\begin{document}

\title{How to Achieve Privacy in Bidirectional\\ Relay Networks}
\author{
  \authorblockN{Rafael F. Wyrembelski
  							and Holger Boche\\[2mm]}
  \authorblockA{Lehrstuhl f\"ur Theoretische Informationstechnik\\
  							Technische Universit\"at M\"unchen, Germany}
\thanks{The authors gratefully acknowledge the support of the TUM Graduate School / Faculty Graduate Center FGC-EI at Technische Universit\"at M\"unchen, Germany. The work of Holger Boche was partly supported by the German Research Foundation (DFG) under Grant BO 1734/25-1.}}
\IEEEoverridecommandlockouts
\maketitle

% ================================================================================================================
% ================================================================================================================
% ================================================================================================================
\begin{abstract}
Recent research developments show that the concept of \textit{bidirectional relaying} significantly improves the performance in wireless networks. This applies to three-node networks, where a half-duplex relay node establishes a bidirectional communication between two other nodes using a decode-and-forward protocol. In this work we consider the scenario when in the broadcast phase the relay transmits additional confidential information to one node, which should be kept as secret as possible from the other, non-intended node. This is the \textit{bidirectional broadcast channel with confidential messages} for which we derive the capacity-equivocation region and the secrecy capacity region. The latter characterizes the communication scenario with perfect secrecy, where the confidential message is completely hidden from the non-legitimated node.
\end{abstract}

% ================================================================================================================
% ================================================================================================================
% ================================================================================================================
\section{Introduction}
\label{sec:introduction}

The use of relays is currently becoming more and more attractive since they have the potential to significantly improve the performance and coverage of wireless networks. Relay communication suffers from the fact that orthogonal resources are needed for transmission and reception. The inherent loss in spectral efficiency can be reduced if bidirectional communication is considered \cite{RankovWittneben07SpectrallyEfficientRelay,Larsson05CodedBidirectionalRelaying}.

Cellular system operators offer for several users different services simultaneously where some of them are subject to secrecy constraints. Due to the nature of the wireless medium, a transmitted signal is received by the intended user but can also be overheard by non-intended users. Consequently, a system design that enables secure communication becomes an important issue especially for confidential information, where non-legitimated receivers should be kept ignorant of it.  

In his seminal work \cite{Wyner75WiretapChannel} Wyner characterized the secure communication problem for a single source-destination link with an eavesdropper, the so-called \textit{wiretap channel}. In \cite{CsiszarKoerner78BroadcastChannelsConfidentialMessages} Csisz\'ar and K\"orner generalized this model and studied the \textit{broadcast channel with confidential messages}. Recently, the secure communication problem gained a lot of attention; for a current survey we refer, for example, to \cite{Liang09InformationTheoreticSecurity}. The multiple access channel with confidential messages is analyzed in \cite{Liang08MACConfidentail}, while \cite{Liu08InterferenceBroadcastConfidential} discusses the interference and broadcast channel. Secure communication in relay broadcast channels is addressed in \cite{Ekrem08SecrecyRelayBroadcast} and in two-way wiretap channels in \cite{He10SecrecyGaussianTwoWayWiretap}.

\begin{figure}
  \centering
  \subfigure[t][MAC phase]{\label{fig:mac_phase}%
    \includegraphics[width=0.22\textwidth]{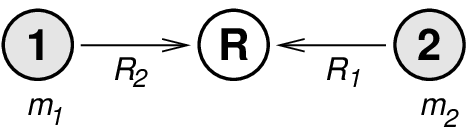}}
  \hspace{0.35cm}
  \subfigure[t][BBC phase]{\label{fig:bc_phase}%
    \includegraphics[width=0.22\textwidth]{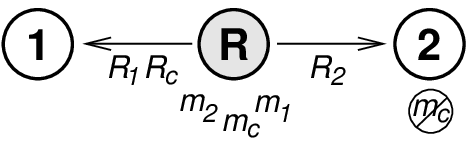}}
  %\vspace{-0.3cm}
  \caption{Decode-and-forward bidirectional relaying. In the initial MAC phase, nodes 1 and 2 transmit their messages $m_1$ and $m_2$ with rates $\Rt$ and $\Ro$ to the relay node. Then, in the BBC phase, the relay forwards the messages $m_1$ and $m_2$ and adds a confidential message $m_c$ for node 1 with rate $\Rc$ to the communication which should be kept as secret as possible from node 2.}
  \label{fig:model}
\end{figure}

We consider \textit{bidirectional relaying} in a three-node network, where a relay node establishes a bidirectional communication between two nodes using a two-phase decode-and-forward protocol as shown in Figure \ref{fig:model}. Here, our main concern is on enabling an additional confidential communication within such a network. This differs from the wiretap scenario where the bidirectional communication itself should be secure from eavesdroppers outside of the wireless network as, for example, studied from a signal processing point of view in \cite{AlSayedSezgin10SecrecyMIMOBBCWiretap,Mukherjee10SecuringTwoWayAnalogNetworkCoding}.

In this work, we concentrate on the broadcast phase, where the relay has successfully decoded the two messages the nodes have sent in the previous multiple access (MAC) phase. The task of the relay is then to transmit both messages and an additional confidential message to one node, which should be kept as secret as possible from the other, non-legitimated node. For decoding, the receiving nodes can exploit the messages they have sent in the previous phase as side information so that this channel differs from the classical broadcast channel with confidential messages and is therefore called \textit{bidirectional broadcast channel (BBC) with confidential messages}.

For the BBC without confidential messages in \cite{Oechtering08BroadcastCapacityRegion,Kim08PerformanceBoundsBidirectional} it is shown that capacity is achieved by a single data stream, which combines both messages based on the network coding idea. Here, we address the problem of realizing additional confidential communication within a network that exploits principles from network coding; hence, the optimal processing is by no means self-evident.\footnote{\textit{Notation:} Discrete random variables are denoted non-italic capital letters and their realizations and ranges by lower case letters and script letters, respectively; $\sP(\cdot)$ denotes the set of all probability distributions and $\An(\cdot)$ the set of (weakly) typical sequences, cf. for example \cite{CoverThomas06ElementsInformationTheory}.}

% ================================================================================================================
% ================================================================================================================
% ================================================================================================================
\section{Bidirectional Broadcast Channel with Confidential Messages}
\label{sec:bbc}

Let $\sX$ and $\sY_i$, $i=1,2$, be finite input and output sets. Then for input and output sequences $x^n\in\sX^n$ and $y_i^n\in\sY_i^n$, $i=1,2$, of length $n$, the discrete memoryless \textit{broadcast channel} is given by $W\tn(y_1^n,y_2^n|x^n):=\prod_{k=1}^nW(y_{1,k},y_{2,k}|x_k)$. Since we do not allow any cooperation between the receiving nodes, it is sufficient to consider the marginal transition probabilities $W_i\tn:=\prod_{k=1}^nW_i(y_{i,k}|x_k)$, $i=1,2$, only.

In this work we consider the standard model with a block code of arbitrary but fixed length $n$. Let $\sM_i:=\{1,...,\Min\}$ be the set of individual messages of node $i$, $i=1,2$, which is also known at the relay node. Further, $\sM_c:=\{1,...,\Mcn\}$ is the set of confidential messages of the relay node. We use the abbreviation $\sM:=\sM_c\times\sM_1\times\sM_2$.

For the bidirectional broadcast (BBC) phase we assume that the relay has successfully decoded the individual messages $m_1\in\sM_1$ from node 1 and $m_2\in\sM_2$ from node 2 that it received in the previous multiple access phase (MAC) phase. Then the relay transmits both individual messages and an additional confidential message $m_c\in\sM_c$ to node 1, which should be kept as secret as possible from node 2.

\begin{definition}
\label{def:code}
An $(n,\Mcn,\Mon,\Mtn)$-\textit{code} for the BBC with confidential messages consists of one (stochastic) encoder at the relay node
\begin{equation*}
	f: \sM_c\times\sM_1\times\sM_2\rightarrow\sX^n
\end{equation*}
and decoders at nodes 1 and 2
\begin{align*}
	g_1&: \sY_1^n\times\sM_1\rightarrow\sM_c\times\sM_2\cup\{0\},\\
	g_2&: \sY_2^n\times\sM_2\rightarrow\sM_1\cup\{0\},
\end{align*}
where the element $0$ in the definition of the decoders plays the role of an erasure symbol and is included for convenience.
\end{definition}

Since randomization may increase secrecy \cite{CsiszarKoerner78BroadcastChannelsConfidentialMessages,Liang09InformationTheoreticSecurity}, we allow  the encoder $f$ to be stochastic. This means it is specified by conditional probabilities $f(x^n|m)$ with $\sum_{x^n\in\sX^n}f(x^n|m)=1$ for each $m=(m_c,m_1,m_2)\in\sM$. Here, $f(x^n|m)$ is the probability that the message $m\in\sM$ is encoded as $x^n\in\sX^n$. 

A code is measured by two performance criteria. First, all transmitted messages have to be successfully decoded, i.e., we want the average probability of a decoding error to be small. In more detail, when the relay has sent the message $m=(m_c,m_1,m_2)$, and nodes 1 and 2 have received $y_1^n$ and $y_2^n$, the decoder at node 1 is in error if $g_1(y_1^n,m_1)\neq(m_c,m_2)$. Accordingly, the decoder at node 2 is in error if $g_2(y_2^n,m_2)\neq m_1$. Then, the average probability of error at node $i$ is given by $\ave_i := \frac{1}{|\sM|}\sum_{m\in\sM}\lambda_i(m)$, $i=1,2$, where $\lambda_1(m)$ denotes the probability that node 1 decodes $(m_c,m_2)$ incorrectly if $m=(m_c,m_1,m_2)$ has been sent, and $\lambda_2(m)$ the probability that node 2 decodes $m_1$ incorrectly.

The second criterion is security. Similarly as in \cite{Wyner75WiretapChannel,CsiszarKoerner78BroadcastChannelsConfidentialMessages} we characterize the secrecy level of the confidential message $m_c\in\sM_c$ at node 2 by the concept of equivocation. The equivocation $H(\rM_c|\rY_2^n,\rM_2)$ describes the uncertainty of node 2 about the confidential message $\rM_c$ having its own message $\rM_2$ and the received sequence $\rY_2^n$ as side information available. Thus, the higher the equivocation, the more ignorant is node 2 about the confidential message.

\begin{definition}
\label{def:achievable}
A rate-equivocation tuple $(\Rc,\Requi,\Ro,\Rt)\in\R_+^4$ is said to be \textit{achievable} for the BBC with confidential messages if for any $\delta>0$ there is an $n(\delta)\in\N$ and a sequence of $(n,\Mcn,\Mon,\Mtn)$-codes such that for all $n\geq n(\delta)$ we have $\frac{\log\Mcn}{n}\geq \Rc-\delta$, $\frac{\log\Mtn}{n}\geq \Ro-\delta$, $\frac{\log\Mon}{n}\geq \Rt-\delta$, and 
\begin{equation}
	\tfrac{1}{n}H(\rM_c|\rY_2^n,\rM_2)\geq\Requi-\delta
	\label{eq:bbc_equivocation}
\end{equation}
while $\ave_1,\ave_2\rightarrow0$ as $n\rightarrow\infty$. The set of all achievable rate-equivocation tuples is the \textit{capacity-equivocation region of the BBC with confidential messages} and is denoted by $\sCbbc$.
\end{definition}

If there is no additional confidential message for the relay to transmit, we have the classical BBC for which the capacity-achieving coding strategies are known \cite{Oechtering08BroadcastCapacityRegion,Kim08PerformanceBoundsBidirectional}.

\begin{theorem}[\cite{Oechtering08BroadcastCapacityRegion,Kim08PerformanceBoundsBidirectional}]
\label{the:bbc}
The capacity region of the BBC is the set of all rate pairs $(\Ro,\Rt)\in\R_+^2$ satisfying
\begin{equation*}
	\Ro\leq I(\rX;\rY_1|\rQ), \quad \Rt\leq I(\rX;\rY_2|\rQ)
\end{equation*}
for random variables $(\rQ,\rX,\rY_1,\rY_2)\in\sQ\times\sX\times\sY_1\times\sY_2$ and joint probability distribution $P_\rQ(q)P_{\rX|\rQ}(x|q)W(y_1,y_2|x)$. The cardinality of the range of $\rQ$ can be bounded by $|\sQ|\leq2$.
\end{theorem}

Now, we focus our attention on the broadcast scenario with a confidential message and present the main result of this work.

\begin{theorem}
\label{the:capacity}
The capacity-equivocation region $\sCbbc$ of the BBC with confidential messages is a closed convex set of those rate-equivocation tuples $(\Rc,\Requi,\Ro,\Rt)\in\R_+^4$ satisfying 
\begin{align*}
	0 \leq \Requi &\leq \Rc, \\
	\Requi &\leq I(\rV;\rY_1|\rU) - I(\rV;\rY_2|\rU), \\
	\Rc + \Ri &\leq I(\rV;\rY_1|\rU) + I(\rU;\rY_i), \quad i=1,2, \\
	\Ri &\leq I(\rU;\rY_i), \quad i=1,2,
\end{align*}
for random variables $(\rU,\rV,\rX,\rY_1,\rY_2)\in\sU\times\sV\times\sX\times\sY_1\times\sY_2$ and joint probability distribution $P_\rU(u)P_{\rV|\rU}(v|u)P_{\rX|\rV}(x|v)W(y_1,y_2|x)$. Moreover, the cardinalities of the ranges of $\rU$ and $\rV$ can be bounded by
\begin{equation*}
	|\sU| \leq |\sX|+3, \qquad |\sV| \leq |\sX|^2+4|\sX|+3.
\end{equation*}
\end{theorem}

\begin{remark}
\label{rem:auxiliary}
While for the BBC without confidential messages the auxiliary random variable $\rQ$ only enables a time-sharing operation and carries no information, cf. Theorem~\ref{the:bbc}, for the BBC with confidential messages we will see that the auxiliary random variable $\rU$ carries the bidirectional information and $\rV$ realizes an additional randomization.
\end{remark}

From Theorem \ref{the:capacity} follows immediately the \textit{secrecy capacity region} $\sCbbc^S$ \textit{of the BBC with confidential messages} which is the set of rate triples $(\Rc,\Ro,\Rt)\in\R_+^3$ such that $(\Rc,\Rc,\Ro,\Rt)\in\sCbbc$.

\begin{corollary}
\label{cor:secrecy}
The secrecy capacity region $\sCbbc^S$ of the BBC with confidential messages is the set of all rate triples $(\Rc,\Ro,\Rt)\in\R_+^3$ satisfying
\begin{align*}
	\Rc &\leq I(\rV;\rY_1|\rU) - I(\rV;\rY_2|\rU), \\
	\Ri &\leq I(\rU;\rY_i), \quad i=1,2, 
\end{align*}
for random variables $(\rU,\rV,\rX,\rY_1,\rY_2)\in\sU\times\sV\times\sX\times\sY_1\times\sY_2$ and joint probability distribution $P_\rU(u)P_{\rV|\rU}(v|u)P_{\rX|\rV}(x|v)W(y_1,y_2|x)$. \endproof
\end{corollary}

The capacity-equivocation region in Theorem \ref{the:capacity} describes the scenario where the confidential message is transmitted with rate $\Rc$ at a certain secrecy level $\Requi$. Thereby, $\Requi$ can be interpreted as the amount of information of the confidential message that can be kept secret from the non-legitimated node. Therefore, Theorem \ref{the:capacity} includes the case where the non-legitimated node has some partial knowledge about the confidential information, namely if $\Rc>\Requi$. The secrecy capacity region in Corollary \ref{cor:secrecy} characterizes the scenario with perfect secrecy which is, of course, the practically more relevant case. Since $\Rc=\Requi$, the confidential message can be kept completely hidden from the non-legitimated node.

% ================================================================================================================
% ================================================================================================================
% ================================================================================================================
\section{Secrecy-Achieving Coding Strategy}
\label{sec:achievability}

In this section, we present a coding strategy that achieves the desired rates with the required secrecy level and therewith prove the achievability part of the corresponding Theorem \ref{the:capacity}.

% ================================================================================================================
\subsection{Codebook Design}
\label{sec:achievability_codebook}

A crucial part is the following Lemma~\ref{lem:codebook} which ensures the existence of a suitable codebook with a specific structure consisting of two layers.

The first layer corresponds to a codebook suitable for the BBC with common messages \cite{Wyrembelski10BidirectionalCommonMessage} which means that this set of codewords enables the relay to transmit (bidirectional) individual messages $m_2'\in\sM_2'$ and $m_1'\in\sM_1'$ to nodes 1 and 2 as well as a common (multicast) message $m_0'\in\sM_0'$ to both nodes.

Then, for each codeword there is a sub-codebook with a product structure similarly as in \cite{CsiszarKoerner78BroadcastChannelsConfidentialMessages} for the classical broadcast channel with confidential messages. The legitimate receiver for the confidential message, i.e., node 1, can decode each codeword regardless to which column and row index it corresponds. But the main idea behind such a codebook design is that the non-legitimated receiver, i.e., node 2, decodes the column index of the transmitted codeword with the maximum rate its channel provides, and therefore is not able to decode the remaining row index \cite{Liang09InformationTheoreticSecurity}.

\begin{lemma}
\label{lem:codebook}
For any $\delta>0$ let $\rU\rightarrow\rX\rightarrow\rY_1\rY_2$ be a Markov chain of random variables and $I(\rX;\rY_1|\rU)>I(\rX;\rY_2|\rU)$. 

\textit{i)} There exists a set of codewords $u_{m'}^n\in\sU^n$, $m'=(m_0',m_1',m_2')\in\sM_0'\times\sM_1'\times\sM_2'=:\sM'$, with
\begin{subequations}
\label{eq:achievability_codebook1}
\begin{align}
	\tfrac{1}{n}\big(\log|\sM_0'|+\log|\sM_2'|\big) &\geq I(\rU;\rY_1)-\delta, \label{eq:achievability_codebook1a} \\
	\tfrac{1}{n}\big(\log|\sM_0'|+\log|\sM_1'|\big) &\geq I(\rU;\rY_2)-\delta, \label{eq:achievability_codebook1b}
\end{align}
\end{subequations}
such that
\begin{subequations}
\label{eq:achievability_codebook2}
\begin{align}
	\frac{1}{|\sM'|}\sum_{m'\in\sM'}\lambda_{m_0',m_2'|m_1'} &\leq \epsilon^{(n)}, \label{eq:achievability_codebook2a} \\
	\frac{1}{|\sM'|}\sum_{m'\in\sM'}\lambda_{m_0',m_1'|m_2'} &\leq \epsilon^{(n)}, \label{eq:achievability_codebook2b}
\end{align}
\end{subequations}
and $\epsilon^{(n)}\rightarrow0$ as $n\rightarrow\infty$. Thereby, $\lambda_{m_0',m_2'|m_1'}$ denotes the probability that node 1 decodes $(m_0',m_2')\in\sM_0'\times\sM_2'$ incorrectly if $m_1'\in\sM_1'$ is given. The error event $\lambda_{m_0',m_1'|m_2'}$ for node 2 is defined accordingly.

\textit{ii)} For each $u_{m'}^n\in\sU^n$ there exists a set of (sub-)codewords $x_{jlm'}^n\in\sX^n$, $j\in\sJ$, $l\in\sL$, $m'\in\sM'$, with
\begin{subequations}
\label{eq:achievability_codebook3}
\begin{align}
	\tfrac{1}{n}\log|\sJ| &\geq I(\rX;\rY_2|\rU)-\delta, \label{eq:achievability_codebook3a}\\
	\tfrac{1}{n}\log|\sL| &\geq I(\rX;\rY_1|\rU) - I(\rX;\rY_2|\rU)-\delta, \label{eq:achievability_codebook3b}
\end{align}
\end{subequations}
such that
\begin{subequations}
\label{eq:achievability_codebook4}
\begin{align}
	\frac{1}{|\sJ||\sL||\sM'|}\sum_{j\in\sJ}\sum_{l\in\sL}\sum_{m'\in\sM'}\lambda_{j,l|m'}&\leq\epsilon^{(n)},\label{eq:achievability_codebook4a} \\
	\frac{1}{|\sJ||\sL||\sM'|}\sum_{j\in\sJ}\sum_{l\in\sL}\sum_{m'\in\sM'}\lambda_{j|l,m'}&\leq\epsilon^{(n)},\label{eq:achievability_codebook4b}
\end{align}
\end{subequations}
and $\epsilon^{(n)}\rightarrow0$ as $n\rightarrow\infty$. Here, $\lambda_{j,l|m'}$ is the probability that node 1 decodes $j\in\sJ$ or $l\in\sL$ incorrectly if $m'\in\sM'$ is known. Similarly, $\lambda_{j|l,m'}$ is the probability that node 2 decodes $j\in\sJ$ incorrectly if $m'\in\sM'$ and $l\in\sL$ are given.

\begin{IEEEproof}[Sketch of Proof]
Since the proof is based on the classical broadcast channel with confidential messages \cite{CsiszarKoerner78BroadcastChannelsConfidentialMessages} and the BBC with common messages \cite{Wyrembelski10BidirectionalCommonMessage} we only sketch the main ideas.

For the first layer we generate $|\sM'|$ codewords $u_{m'}^n\in\sU^n$ according to the distribution $P_{\rU^n}(u^n)=\prod_{k=1}^nP_\rU(u_k)$ and use (weakly) typical sets $\An(\rU,\rY_i)$, $i=1,2$, for decoding at the receivers. Then, using random coding arguments, for the BBC with common messages we know from \cite{Wyrembelski10BidirectionalCommonMessage} that (\ref{eq:achievability_codebook2}) is satisfied if (\ref{eq:achievability_codebook1}) is fulfilled proving the first part.

To prove the second assertion, for each $u_{m'}^n\in\sU^n$ we generate $|\sJ||\sL|$ codewords $x_{jlm'}^n\in\sX^n$ according to $P_{\rX^n|\rU^n}(x^n|u^n)=\prod_{k=1}^nP_{\rX|\rU}(x_k|u_k)$ and use typical sets $\An(\rU,\rX,\rY_i)$, $i=1,2$, for decoding at the receivers. We note that the structure of the sub-codewords is exactly the same as for the classical broadcast channel with confidential messages \cite{CsiszarKoerner78BroadcastChannelsConfidentialMessages,Liang09InformationTheoreticSecurity}, where the latter assumes the average error criterion and uses random coding arguments as we do. Following the proof it is easy to show that (\ref{eq:achievability_codebook4}) is satisfied if (\ref{eq:achievability_codebook3}) is fulfilled proving the second part.
\end{IEEEproof}
\end{lemma}

% ================================================================================================================
\subsection{Achievable Rate-Equivocation Region}
\label{sec:achievability_encoder}

Next, we use the codebook from Lemma \ref{lem:codebook} to construct suitable encoder and decoders for the BBC with confidential messages.

\begin{lemma}
\label{lem:region}
Let $\rU\rightarrow\rX\rightarrow\rY_1\rY_2$ and $I(\rX;\rY_1|\rU)>I(\rX;\rY_2|\rU)$. Using the codebook from Lemma \ref{lem:codebook} all rate-equivocation tuples $(\Rc,\Requi,\Ro,\Rt)\in\R_+^4$ satisfying
\begin{subequations}
\label{eq:achievability_region}
\begin{align}
	0 \leq \Requi &= I(\rX;\rY_1|\rU) - I(\rX;\rY_2|\rU) \leq \Rc, \label{eq:achievability_regiona} \\
	\Rc+\Ri &\leq I(\rX;\rY_1|\rU) + I(\rU;\rY_i), \quad i=1,2, \label{eq:achievability_regionb} \\
	\Ri &\leq I(\rU;\rY_i), \quad i=1,2, \label{eq:achievability_regionc}
\end{align}
\end{subequations}
are achievable for the BBC with confidential messages.
\begin{IEEEproof}
For given rate-equivocation tuple $(\Rc,\Requi,\Ro,\Rt)\in\R_+^4$ satisfying (\ref{eq:achievability_regiona})-(\ref{eq:achievability_regionc}) we have to construct message sets, encoders, and decoders with
\begin{subequations}
\label{eq:achievability_rates}
\begin{align}
	\tfrac{1}{n}\log|\sM_c|\geq\Rc-\delta, \label{eq:achievability_ratesc} \\
	\tfrac{1}{n}\log|\sM_2|\geq\Ro-\delta, \label{eq:achievability_rates1} \\
	\tfrac{1}{n}\log|\sM_1|\geq\Rt-\delta, \label{eq:achievability_rates2}
\end{align}
\end{subequations}
and further, cf. also (\ref{eq:bbc_equivocation}),
\begin{equation}
	\tfrac{1}{n}H(\rM_c|\rY_2^n,\rM_2) \geq I(\rX;\rY_1|\rU) - I(\rX;\rY_2|\rU)-\delta.
	\label{eq:achievability_equivocation}
\end{equation}
The following construction is mainly based on the one for the classical broadcast channel with confidential messages \cite{CsiszarKoerner78BroadcastChannelsConfidentialMessages}. Thereby, we have to distinguish between two cases as visualized in Figures \ref{fig:enc1} and \ref{fig:enc2}.

\begin{figure}
  \centering
    \scalebox{0.7}{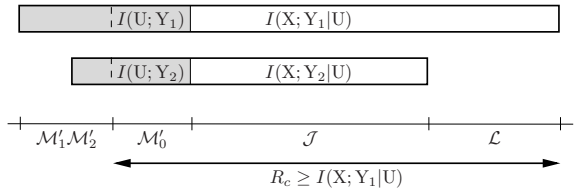}
  \vspace{-0.25cm}
  \caption{The two bars visualize the available resources of both links. Each one is split up into two parts: one designated for the bidirectional communication (gray) and one for the confidential message (white). Since $\Rc\geq I(\rX;\rY_1|U)$, some resources of the bidirectional communication have to be spent for the confidential message as well (realized by a common message).}
  \label{fig:enc1}
\end{figure}

If $\Rc\geq I(\rX;\rY_1|\rU)$, cf. Figure \ref{fig:enc1}, we construct the set of confidential messages as
\begin{equation*}
	\sM_c := \sJ\times\sL\times\sM_0'
\end{equation*}
where $\sJ$ and $\sL$ are chosen as in Lemma \ref{lem:codebook} and $\sM_0'$ is an arbitrary set of common messages such that (\ref{eq:achievability_ratesc}) is satisfied. The sets $\sM_1=\sM_1'$ and $\sM_2=\sM_2'$ are arbitrary such that (\ref{eq:achievability_rates1})-(\ref{eq:achievability_rates2}) hold. Finally, we define the deterministic encoder $f$ that maps the confidential message $(j,l,m_0')\in\sM_c$ and the individual messages $m_i\in\sM_i$, $i=1,2$, into the codeword $x_{jlm'}^n\in\sX^n$ with $m'=(m_0',m_1',m_2')$ and $m_i'=m_i$, $i=1,2$.

\begin{remark}
\label{rem:common}
Since $\Rc\geq I(\rX;\rY_1|\rU)$, a part of the confidential message must be transmitted as a common message. It is not possible to simply "add" the remaining part to the individual message for node 1, since this would require that this part of the confidential message is already available a priori as side information at node 2.
\end{remark}

If $\Rc<I(\rX;\rY_1|\rU)$, cf. Figure \ref{fig:enc2}, we set $\sM_c:=\sK\times\sL$ where $\sK$ is an arbitrary set such that (\ref{eq:achievability_ratesc}) holds. Further, we define a mapping $h:\sJ\rightarrow\sK$ that partitions $\sJ$ into subsets of "nearly equal size" \cite{CsiszarKoerner78BroadcastChannelsConfidentialMessages}, which means
\begin{equation*}
	|h^{-1}(k)| \leq 2|h^{-1}(k')|, \quad\text{for all }k,k'\in\sK.
\end{equation*}
Moreover, since $\Rc<I(\rX;\rY_1|\rU)$, there is no need for a set of common messages so that $\sM_0'=\emptyset$. The sets $\sM_1=\sM_1'$ and $\sM_2=\sM_2'$ are arbitrary such that (\ref{eq:achievability_rates1})-(\ref{eq:achievability_rates2}) hold. Finally, we define the stochastic encoder $f$ that maps the confidential message $(k,l)\in\sM_c$ and the individual messages $m_i\in\sM_i$, $i=1,2$, into the codeword $x_{jlm'}^n\in\sX^n$ with $m'=(0,m_1',m_2')$, where $j$ is uniformly drawn from the set $h^{-1}(k)\subset\sJ$ and $m_i'=m_i$, $i=1,2$.

\begin{remark}
\label{rem:stochastic}
This time, the set $\sJ$ is not needed in total for the confidential message. However, to force the non-legitimated receiver, i.e., node 2, to decode at its maximum rate, we define a stochastic encoder that spreads the confidential messages over the whole set $\sJ$.
\end{remark}

Up to now we defined message sets and the encoder. In both cases the decoders are immediately determined by Lemma \ref{lem:codebook}. 

\begin{figure}
  \centering
    \scalebox{0.7}{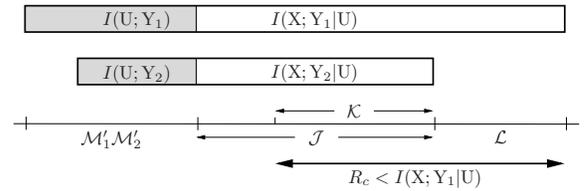}
  \vspace{-0.25cm}
  \caption{Since $\Rc< I(\rX;\rY_1|U)$, there are more resources for the confidential message available than needed. This allows the relay to enable a stochastic coding strategy which exploits all the available resources by introducing a mapping from $\sJ$ to $\sK$. }
  \label{fig:enc2}
\end{figure}

To complete the proof it remains to show that the secrecy level at node 2 fulfills (\ref{eq:achievability_equivocation}). Proceeding exactly as in \cite{CsiszarKoerner78BroadcastChannelsConfidentialMessages} we define the random variable $\rX^n$ with codewords $x_{jlm'}^n\in\sX^n$ as realizations and $\rM'=(\rM_0',\rM_1',\rM_2')$ as the third coordinate of the realization of $\rX^n$. Then get for the equivocation
\begin{align}
	\begin{split}
	&H(\rM_c|\rY_2^n,\rM_2) \geq H(\rX^n|\rM') + H(\rY_2^n|\rX^n) \\
		&\quad\qquad\qquad - H(\rX^n|\rM_c,\rM',\rY_2^n) - H(\rY_2^n|\rM').
		\label{eq:achievability_equibound}
		\end{split}
\end{align}
Next, we bound all terms in (\ref{eq:achievability_equibound}) separately. We start with the first term and observe for given $\rM'=m'$ that $\rX^n$ has $|\sJ||\sL|$ possible values. Since $\rX^n$ is independently and uniformly distributed, we have $H(\rX^n|\rM')=\log|\sJ|+\log|\sL|$. With the definition of $\sJ$ and $\sL$, cf. (\ref{eq:achievability_codebook3}), we obtain
\begin{equation}
	\tfrac{1}{n}H(\rX^n|\rM')\rightarrow I(\rX;\rY_1|\rU).
	\label{eq:achievability_equibound1}
\end{equation}
For the second term in (\ref{eq:achievability_equibound}) we have
\begin{equation}
	\tfrac{1}{n}H(\rY_2^n|\rX^n) \rightarrow H(\rY_2|\rX)
\label{eq:achievability_equibound2}
\end{equation}
as $n\rightarrow\infty$ by the weak law of large numbers.
If $\Rc\geq I(\rX;\rY_1|\rU)$, the third term in (\ref{eq:achievability_equibound}) vanishes. If $\Rc< I(\rX;\rY_1|\rU)$, we define $\varphi(k,l,m',y_2^n) := x_{klm'}^n$ if $(u_{m'}^n,x_{jlm'}^n,y_2^n)\in\An(\rU,\rX,\rY_2)$, $h(j)=k$, and arbitrary otherwise. Then we have $\prob\{\rX^n\neq\varphi(\rM_c,\rM',\rY_2^n)\}\leq\epsilon^{(n)}$ and therefore, by Fano's lemma, cf. also \cite{CsiszarKoerner78BroadcastChannelsConfidentialMessages,Liang09InformationTheoreticSecurity},
\begin{equation}
	\tfrac{1}{n}H(\rX^n|\rM_c,\rM',\rY_2^n)\rightarrow0
	\label{eq:achievability_equibound3}
\end{equation}
as $n\rightarrow\infty$.
For the last term in (\ref{eq:achievability_equibound}) we define $\hat{y}_2^n := y_2^n$ if $(u_{m'}^n,y_2^n)\in\An(\rU,\rY_2)$ and arbitrary otherwise so that
\begin{equation*}
	H(\rY_2^n|\rM') \leq H(\rY_2^n|\hat{\rY}_2^n) + H(\hat{\rY}_2^n|\rM').
\end{equation*}
For the first term we have $\prob\{\rY_2^n\neq\hat{\rY}_2^n\}\leq\epsilon^{(n)}$ by Fano's lemma, cf. \cite{CsiszarKoerner78BroadcastChannelsConfidentialMessages,Liang09InformationTheoreticSecurity}, so that it is negligible. Moreover, following \cite{CsiszarKoerner78BroadcastChannelsConfidentialMessages,Liang09InformationTheoreticSecurity} it is easy to show that for the second term we have
\begin{equation}
	\tfrac{1}{n}H(\hat{\rY}_2^n|\rM') \rightarrow H(\rY_2|\rU)
	\label{eq:achievability_equibound4}
\end{equation}
which follows from the definition of the decoding sets $\An(\rU,\rY_2)$ and the fact that the codewords are uniformly distributed.

Finally, by substituting (\ref{eq:achievability_equibound1})-(\ref{eq:achievability_equibound4}) into (\ref{eq:achievability_equibound}) we obtain (\ref{eq:achievability_equivocation}) which establishes the desired secrecy level at node 2 and therewith proves the lemma. 
\end{IEEEproof}
\end{lemma}

% ================================================================================================================
\subsection{Randomization and Convexity}
\label{sec:achievability_randomization}

Here, we complete the proof of achievability of Theorem~\ref{the:capacity}. Since the argumentation is the same as for the classical broadcast channel with confidential messages \cite{CsiszarKoerner78BroadcastChannelsConfidentialMessages}, we only sketch the main ideas.

To obtain the whole region of Theorem \ref{the:capacity}, we proceed exactly as in \cite{CsiszarKoerner78BroadcastChannelsConfidentialMessages} and introduce an auxiliary channel that enables an additional randomization.

\begin{lemma}
\label{lem:prefix}
Let $\rU\rightarrow\rV\rightarrow\rX\rightarrow\rY_1\rY_2$ and $I(\rV;\rY_1|\rU)>I(\rV;\rY_2|\rU)$. Then all rate-equivocation tuples $(\Rc,\Requi,\Ro,\Rt)\in\R_+^4$ satisfying
\begin{subequations}
\label{eq:achievability_prefix}
\begin{align}
	0 \leq \Requi &\leq I(\rV;\rY_1|\rU) - I(\rV;\rY_2|\rU) \leq \Rc, \label{eq:achievability_prefix1} \\
	\Rc + \Ri &\leq I(\rV;\rY_1|\rU) + I(\rU;\rY_i), \quad i=1,2, \label{eq:achievability_prefix2} \\
	\Ri &\leq I(\rU,\rY_i), \quad i=1,2, \label{eq:achievability_prefix3}
\end{align}
\end{subequations}
are achievable for the BBC with confidential messages. The corresponding rate region is denoted by $\sR$.
\begin{IEEEproof}[Sketch of Proof]
The prefixing realized by the random variable $\rV$ is exactly the same as in \cite[Lemma 4]{CsiszarKoerner78BroadcastChannelsConfidentialMessages}.

Moreover, it is obvious that if the rate-equivocation tuple $(\Rc,\Requi,\Ro,\Rt)$ is achievable, than each rate-equivocation tuple $(\Rc,\Requi',\Ro,\Rt)$ with $0\leq\Requi'\leq\Requi$ is also achievable. Consequently, we can further replace the equality in (\ref{eq:achievability_regiona}) by an inequality in (\ref{eq:achievability_prefix1}).
\end{IEEEproof}
\end{lemma}

\begin{lemma}
\label{lem:convex}
The rate region $\sR$ is convex.
\begin{IEEEproof}[Sketch of Proof]
Exactly as in \cite[Lemma 5]{CsiszarKoerner78BroadcastChannelsConfidentialMessages} it is easy to show that any linear combination of two rate tuples in $\sR$ is contained in $\sR$ which proves the convexity.
\end{IEEEproof}
\end{lemma}

It remains to show that $\sR$ describes the same rate region as the one specified by Theorem \ref{the:capacity}.

\begin{lemma}
\label{lem:equal}
The rate region $\sR$ equals the capacity region $\sCbbc$ of the BBC with confidential messages.
\begin{IEEEproof}
It is obvious that $\sR\subseteq\sCbbc$ holds. To show the reversed inclusion, i.e., $\sCbbc\subseteq\sR$, let $(\Rc,\Requi,\Ro,\Rt)\in\sCbbc$ be any rate-equivocation tuple. For this, we construct as in \cite{CsiszarKoerner78BroadcastChannelsConfidentialMessages} the maximal achievable confidential and equivocation rates that are possible for given individual rates $\Ro$ and $\Rt$ as
\begin{align*}
	\Rc^* &:= I(\rV;\rY_1|\rU) + \min\big\{I(\rU;\rY_1)\!-\!\Ro,I(\rU;\rY_2)\!-\!\Rt\big\}, \\
	\Requi^* &:= I(\rV;\rY_1|\rU) - I(\rV;\rY_2|\rU).
\end{align*}
Then we have $\Requi\leq\Requi^*$, $\Requi^*\leq\Rc\leq\Rc^*$, and therewith also $(\Rc^*,\Requi^*,\Ro,\Rt)\in\sR$. Now, from the definition of $\sR$ follows that the rate-equivocation tuples $(\Rc^*,\Requi^*,\Ro,\Rt)$, $(\Rc^*,0,\Ro,\Rt)$, and $(0,0,\Ro,\Rt)$ belong to $\sR$ as well. Finally, from the convexity of $\sR$, cf. Lemma \ref{lem:convex}, follows that $(\Rc,\Requi,\Ro,\Rt)\in\sR$ which proves the lemma.
\end{IEEEproof}
\end{lemma}

To complete the proof of achievability it remains to bound the cardinalities of the ranges of $\rU$ and $\rV$. Since the bounds of the cardinalities depend only the structure of the random variables, the result follows immediately from \cite[Appendix]{CsiszarKoerner78BroadcastChannelsConfidentialMessages} where the same bounds are established for the classical broadcast channel with confidential messages.

% ================================================================================================================
\subsection{Weak Converse}
\label{sec:achievability_optimality}

Already the coding strategy indicates that, basically, ideas from the BBC \cite{Oechtering08BroadcastCapacityRegion,Wyrembelski10BidirectionalCommonMessage} and from the classical broadcast channel with confidential messages \cite{CsiszarKoerner78BroadcastChannelsConfidentialMessages} are exploited. 
Based on this observation it is straightforward to establish the weak converse for the BBC with confidential messages by extending the converse of the classical broadcast channel with confidential messages \cite{CsiszarKoerner78BroadcastChannelsConfidentialMessages} using standard arguments for the BBC \cite{Oechtering08BroadcastCapacityRegion,Wyrembelski10BidirectionalCommonMessage}.

% ================================================================================================================
% ================================================================================================================
% ================================================================================================================
\section{Discussion}
\label{sec:conclusion}

In this work, our focus was on privacy in bidirectional relay networks, where additionally to the two bidirectional messages the relay node transmits a confidential message to one of the nodes, which should be kept as secret as possible from the other, non-legitimated node. For this scenario we characterized the corresponding capacity-equivocation and secrecy capacity regions in detail. This scenario is completely different from the bidirectional broadcast wiretap channel, where the bidirectional communication itself should be kept secret from eavesdroppers outside of the bidirectional relay network \cite{AlSayedSezgin10SecrecyMIMOBBCWiretap,Mukherjee10SecuringTwoWayAnalogNetworkCoding}. This is an interesting and important topic for itself.

% ================================================================================================================
% ================================================================================================================
% ================================================================================================================
\bibliographystyle{IEEEtran}
\bibliography{D:/Literatur/bibliography}

\end{document}